\documentclass[prl,twocolumn,aps,superscriptaddress,showpacs,floatfix]{revtex4}
\usepackage{amsmath}
\usepackage{graphicx}
\begin{document}
\title{Quantum Tera-Hertz electrodynamics in Layered Superconductors}
\author{Sergey Savel'ev}
\affiliation{Frontier Research System, The Institute of Physical
and Chemical Research (RIKEN), Wako-shi, Saitama, 351-0198, Japan}
\affiliation{Department of Physics, Loughborough University,
Loughborough LE11 3TU, United Kingdom}
\author{A.L. Rakhmanov}
\affiliation{Frontier Research System, The Institute of Physical
and Chemical Research (RIKEN), Wako-shi, Saitama, 351-0198, Japan}
\affiliation{Institute for Theoretical and Applied Electrodynamics
RAS, 125412 Moscow, Russia}
\author{Franco Nori}
\affiliation{Frontier Research System, The Institute of Physical
and Chemical Research (RIKEN), Wako-shi, Saitama, 351-0198, Japan}
\affiliation{Department of Physics and MCTP, University of
Michigan, Ann Arbor, MI 48109-1040, USA}
\date{\today}
\begin{abstract}
In close analogy to quantum electrodynamics, we derive a quantum
field theory of Josephson plasma waves (JPWs) in layered
superconductors (LSCs), which describes two types of interacting
JPW bosonic quanta: one massive and the other almost-massless. We
also calculate the amplitude of their decay and scattering. We
propose a mechanism of enhancement of macroscopic quantum
tunneling (MQT) in stacks of intrinsic Josephson-junctions
(SIJJs). Due to the long-range interactions between many junctions
in the LSCs, the calculated MQT escape rate $\Gamma$ has a very
nonlinear dependence on the number of junctions in the stack. This
allows to quantitatively describe striking recent experiments in
Bi2212 stacks.
\end{abstract}
 \maketitle

The recent surge of interest on Stacks of Intrinsic Josephson
Junctions (SIJJs) is partly motivated by the desire to develop THz
devices, including emitters~\cite{emit,cher}, filters~\cite{filt},
detectors~\cite{detect}, and nonlinear devices~\cite{natphys}.
Macroscopic quantum tunnelling (MQT) has been, until recently,
considered to be negligible in high-T$_c$ superconductors due to
the d-wave symmetry of the order parameter. Recent unexpected
experimental evidence~\cite{MQT,ustMQT} of MQT in layered
superconductors (LSCs) could open a new avenue for the
applicability of SIJJs in quantum electronics~\cite{martinis}.
This requires a quantum theory for SIJJs capable of describing
quantitatively this new stream of remarkable experimental data.
In contrast to a single Josephson junction, SIJJs are strongly
coupled along the direction perpendicular to the layers. This is
because the thickness of these layers is of the order of a few nm,
which is much smaller than the magnetic penetration length. This
results in a profoundly nonlocal electrodynamics~\cite{cher} that
strongly affects the quantum fluctuations in SIJJs.

Using a general Lagrangian approach, we derive the quantum
electrodynamics of JPWs, which describes two interacting quantum
fields. We analyze the first-order Feynman diagrams for: {\it (i)}
the decay of a quantum JPW propagating along the layers and, {\it
(ii)} JPW-JPW scattering.
Employing the quantum statistics of these plasmons, we calculate
the average energy of the JPWs as a function of temperature, and
find it to be much higher than for the same number of
non-interacting junctions.
Using this general approach, we develop a quantitative theory of
the MQT in SIJJs. For example, we derive the MQT escape rate,
$\Gamma$, which is strongly non-linear with respect to the number
of superconducting layers, $N$, and changes to $\Gamma\propto N$
when $N$ exceeds a certain critical value $N_c$. Our results are
in a good quantitative agreement with recent very exciting
experiments~\cite{ustMQT}.

{\it Quantum theory for layered superconductors.---} The
electrodynamics of SIJJs can be described by the Lagrangian:
\begin{eqnarray} \label{l}
{\cal L}\;=\;\sum_n\int
dx\Biggl\{\frac{1}{2}\;\dot{\varphi_n}^2+\frac{1}{2\gamma^2}\dot{p_n}^2-
\frac{1}{2}(\partial_x\varphi_n)^2-\frac{1}{2}(\partial_yp_n)^2
\nonumber
\\
-\;\frac{1}{2}p_n^2 +\cos\varphi_n
+\frac{1}{2}\left(\partial_xp_n\partial_y\varphi_n+\partial_yp_n\partial_x\varphi_n
\right)\Biggr\}\;,
\end{eqnarray}
where $\varphi_n\equiv\chi_{n+1}-\chi_n-2\pi s A_y^{(n)}/\Phi_0$
is the gauge-invariant interlayer phase difference, and
$p_n\equiv(s/\lambda_{ab})\partial_x\chi_n-2\pi\gamma s
A_{x}^{(n)}/\Phi_0$ is the normalized superconducting momentum in
the $n$th layer. Here, we introduce the phase $\chi_n$ of the
order parameter, the interlayer distance $s$, the in-plane
$\lambda_{ab}$ and out-of-plane $\lambda_c$ penetration depths,
the anisotropy parameter $\gamma=\lambda_c/\lambda_{ab}$, flux
quantum $\Phi_0$, and vector potential $\vec A$. The in-plane
coordinate $x$ is normalized by $\lambda_c$; the time $t$ is
normalized by $1/\omega_J$, where the plasma frequency is
$\omega_J$; also, $\partial_x=\partial/\partial x$, $\partial_y
f_n=\lambda_{ab}(f_{n+1}-f_n)/s$, and $\dot{\;}=\partial/\partial
t$. We choose the $z$ axis pointed along the magnetic field.
Varying the action ${\cal S}=\int\! dt\; {\cal L}$ produces the
dynamical equations
\begin{eqnarray}\label{din}
\ddot{\varphi}_n-\partial_x^2\varphi_n+\sin\varphi_n+\partial_x\partial_y
p_n=0, \nonumber \\
\frac{1}{\gamma^2}\ddot{p}_n-\partial^2_{y}p_n+p_n+\partial_x\partial_y\varphi_n=0,
\end{eqnarray}
which reduces to the usual coupled sine-Gordon
equations~\cite{equat} for $\gamma^2\gg 1$. Note that a Lagrangian
approach for SIJJs can be formulated only for two interacting
fields $\varphi$ and $p$, but not for $\varphi$ alone. This
because of the 2D nature of the vector potential in SIJJs. So
particles with two types of polarization can propagate. For a 1D
Josephson junction, only one polarization is enough.

Linearizing Eqs.~(\ref{din}) results in the spectrum
$\omega^2=1+k_x^2/(1+k_y^2)$ of the classical JPWs
 in the continuous limit (i.e.,
$k_ys\ll 1$) and $\gamma^2\gg 1$. Here, $k_x$ and $k_y$ are the
wave vectors (momentums in the quantum description; here,
$\hbar=1$) of the JPWs. In order to quantize the JPWs we introduce
the Hamiltonian, ${\cal H}=\sum_n\int
dx(\Pi_{\varphi_n}\varphi_n+\Pi_{p_n} p_n) - {\cal L}$, with the
momenta $\Pi_{\varphi_n}$ and $\Pi_{p_n}$ of the $\varphi_n$ and
$p_n$ fields, and require the standard commutation relations
$[\Pi_{\varphi_n}(x),\varphi_{n'}(x')]=-i\delta(x-x')\delta_{n,n'}$,
$[\Pi_{p_n}(x),p_{n'}(x')]=-i\delta(x-x')\delta_{n,n'}$ (all
others commutators are zero), 
where $\delta$ is either a delta function or Kronecker symbol.
Expanding $\cos\varphi_n= 1-\varphi_n^2/2+\varphi_n^4/24-...\;$,
we can write ${\cal H}={\cal H}_0+{\cal H}_{\rm an}$, where we
include terms up to $\varphi_n^2$ in ${\cal H}_0$, and the
anharmonic terms in ${\cal H}_{\rm an}$. Diagonalizing ${\cal
H}_0$, we obtain the Hamiltonian for the Bosonic free fields $a$
and $b$:
${\cal H}_0=\sum_{k_y}\int (dk_x/2\pi)
\left\{\varepsilon_{a}({\vec k})\; a^{+}a\; +\;\varepsilon_b({\vec
k})\; b^+b)\right\}$,  
where the energy of the
quasiparticles are
\begin{equation}\label{eab}
\varepsilon_{a}({\vec
k})=\left(1+\frac{k_x^2}{1+k_y^2}\right)^{1/2}, \ \ \
\varepsilon_{b}({\vec
k})=\frac{1}{\gamma}\frac{|k_xk_y|}{\sqrt{k_y^2+1}}
\end{equation}
up to $1/\gamma^2$, for $\gamma\gg 1$. The energy
$\varepsilon_{a}({\vec k})$ coincides with the frequency
$\omega(k_x,k_y)$ for {\it classical} JPWs, while the quantum
bosonic field $b$ corresponds to the gapless branch of the
excitations. The original fields $\varphi_n$, $p_n$ in
Eq.~(\ref{l}) are related to the free Bosonic fields $a$ and $b$
by
\begin{eqnarray}\label{phiab}
\varphi\approx \frac{a^++a}{\sqrt{2\varepsilon_a}}+\frac{{\cal
Z}^2}{\gamma^2}\frac{b^++b}{\sqrt{2\varepsilon_b}},\ \
p\approx{\cal Z}\left(
\frac{a^++a}{\sqrt{2\varepsilon_a}}-\frac{b^++b}{\sqrt{2\varepsilon_b}}\right),
\end{eqnarray}
where ${\cal Z}=k_xk_y/(k_y^2+1)$. The case of a {\it single}
Josephson junction corresponds to $k_y=0$ resulting in a {\it
single} field, $a$.

{\it Thermodynamics of quantum JPWs.---} Finite temperatures, $T$,
excite both $a$ and $b$ quasiparticles providing contributions of
the JPWs to the thermodynamical quantities. The thermal
equilibrium internal energy $E(T)=\sum_{k_y} \int
dk_x/(2\pi)[\varepsilon_a\, n_a+\varepsilon_b\, n_b]$ of the
system can be calculated using the usual Bosonic distributions
$n_{a,b}=1/[\exp(\varepsilon_{a,b}/T)-1]$. The calculated
dependence of $E(T)$ for SIJJs and, for comparison, for an
``equivalent'' stack of non-interacting Josephson junctions is
shown in Fig.~1a. This clearly shows that the thermodynamic
energies are significantly different for these systems, especially
at low temperatures. Thus, finite temperatures easily thermally
excite JPWs in layered superconductors, compared with the case of
non-interacting junctions. The main origin of this enhancement is
the suppression of both excitation energies
$\varepsilon_{a,b}(k_x,k_y)$ when increasing $k_y$, which is
associated with a stronger interlayer interaction. Other
thermodynamic quantities, e.g., heat capacity, can be easily
calculated using the standard
expressions.

{\it Interaction of Bosonic fields.---} The interaction between
the $a$ and $b$ fields, including the self-interaction, occurs due
to the anharmonic terms in ${\cal H}_{\rm an}\approx
(-1/24)\sum_n\int\! dx\;\varphi_n^4\;$+..., where $\varphi$ is
given in~\eqref{phiab}. Here we consider the dominant first-order
perturbation terms. Using the interaction
representation,
we obtain the amplitude $S_{\rm if} = 2\pi i\langle f|{\cal
H}_{\rm an}|i\rangle\delta(\varepsilon_i-\varepsilon_f)$ for the
transition from the initial state $|i\rangle$ to a final state
$|f\rangle$, where $\varepsilon_{i,f}$ are the energies of the
initial and final states. To first-order approximation in
$1/\gamma^2$, a decay (see Fig.~1b) of an $a$-JPW can occur in two
channels: either $3a$ or $2a+b$. The amplitude, $S_{\rm decay}$,
of the decay of the quantum $a$-JPW propagating along the
$x$-axis, i.e., along the layers, ${\vec k_1}=(k_{x1},\;0)$, is
determined by
\begin{eqnarray}\label{decay}
S_{\rm decay}=\int\frac{i\prod d^2{\vec k}_l}{2^8\,3\,\pi^5}
\Biggl\{\frac{\delta\!\left(\varepsilon_a({\vec
k}_1)-\sum\varepsilon_a({\vec
k}_l)\right)}{\sqrt{\varepsilon_a({\vec
k}_1)\prod\varepsilon_a({\vec k}_l)}}\ \ \ \ \ \ \ \ \ \ \ \ \ \
\\+
\frac{\delta\!\left(\varepsilon_a({\vec k}_1)-\varepsilon_a({\vec
k}_2)-\varepsilon_a({\vec k}_3)-\varepsilon_b({\vec
k}_4)\right)}{(4\gamma^2/9)\; {\cal Z}^2({\vec
k}_4)\sqrt{\varepsilon_a({\vec k}_1)\varepsilon_a({\vec
k}_2)\varepsilon_a({\vec k}_3)\varepsilon_b({\vec
k}_4)}}\Biggr\}\;\delta\!\left({\vec k}_1-\sum{\vec
k}_l\right)\nonumber
\end{eqnarray}
where the sums and products are performed over the final states
with momenta ${\vec k}_l$. Eq.~(\ref{decay}) predicts the
probability $|S_{\rm decay}|^2$ to create JPWs propagating
perpendicular to the layers by an $a$-JPW quantum  propagating
along the layers. Using Eqs.~\eqref{eab} and \eqref{decay}, one
can conclude that the amplitude $S_{\rm decay}$ diverges for large
$k_{yl}$, when resonance conditions [$\varepsilon_a({\vec k}_1)=2$
or 3] are fulfilled. For the former case (in dimensional units,
$\varepsilon_a({\vec k}_1)=2\hbar\omega_J$) the $a$-JPWs create
$a$-$b$-JPW pairs, while at $\varepsilon_a=3\hbar\omega_J$ the
$2a$-excitations diverge. Indeed, due to the $\varphi^4$ nonlinear
interaction, a particle can only create two more additional
particles, which could be either $2a$ or $a+b$. The first process
has a threshold $2\hbar\omega_J$ (similar to the $2mc^2$ rest
energy threshold for $e^-+e^+$ pair creation in QED), while the
second one has a $\hbar\omega_J$ energy threshold due to the
gapless nature of the $b$ particles. Figure 1d shows the
calculated probability, $|S_{\rm decay}|^2$, of decay of a
JPW-$a$-quantum versus the energy $\varepsilon_a$ of the initial
$a$-quantum. Both resonance peaks are clearly seen.

We can similarly analyze the scattering of $a$-JPWs. The diagrams
in Fig.~1c show two input particles as the initial state
$|i\rangle$, corresponding to particles ``1'' and ``2'', while the
final state $|f\rangle$ contains free particles ``3'' and ``4''.
These diagrams do not diverge for any input particle momentum.
However, the scattering probability enormously increases for large
transverse momentum transfer ($k_{y1}-k_{y3}$), if the energies of
the initial particles are close to $\hbar\omega_J$. This can occur
either for low $k_x$ or large $k_y$ of the particles 1 and 2. The
decay and scattering resonances occur due to the unusual
anisotropic spectrum of the JPWs, i.e.,
$\varepsilon_a(k_x,k_y\rightarrow\infty)=1$ and
$\varepsilon_b(k_x,k_y\rightarrow\infty)=k_x/\gamma$.

\begin{figure}[!htp]
\begin{center}\label{term}
\includegraphics*[width=8.7cm]{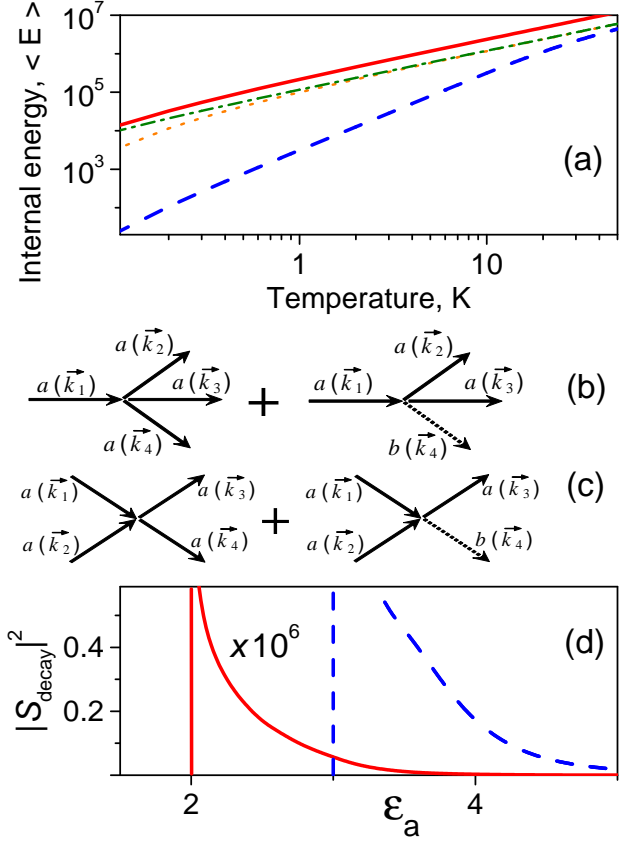}
\caption{(Color online)(a) Thermal energy $\langle E\rangle(T)$ of
quantum JPWs versus temperature $T$ for SIJJs (top red solid
line). The energy $\langle E\rangle$ is normalized by
$\hbar\omega_J$ and sample volume. For comparison, the bottom
dashed blue line shows $\langle E\rangle(T)$ for the {\it same}
number of {\it non}-interacting Josephson junctions and with the
{\it same} parameters. The dashed-dotted green and the dotted
orange lines correspond to the contributions of the $a$ and $b$
fields to the energy, respectively. The chosen parameters are
standard for Bi2212: $s=15$~\AA, $\lambda_{ab}=2000$~\AA,
$\gamma=600$, and $\omega_J/2\pi=150$~GHz. (b,c) Feynmann diagrams
for decay (b) and scattering (c) of JPWs quanta due to anharmonic
interactions. (d) The probability per unit time, $|S_{\rm
decay}|^2$, for an $a$-JPWs to decay, versus energy
$\varepsilon_a({\vec k}_1)$ of the initial particle, for
$\gamma=300$. Both axes are in dimensionless units. The blue
dashed line in (d) corresponds to the $a\rightarrow 3a$ channel
pair-production and the red solid line to the $a\rightarrow 2a+b$
channel. }
\end{center}
\end{figure}

{\it Enhancement of macroscopic quantum tunneling.---} Now we
apply our theory to interpret very recent
experiments~\cite{ustMQT} on MQT in Bi2212. To observe MQT, an
external current $J$, close to the critical value $J_c$, was
applied~\cite{ustMQT}. This produces an additional contribution
$j\varphi_n$ in the Lagrangian~\eqref{l}. When tunneling occurs,
the phase difference in a junction changes from 0 to $2\pi$, which
can be interpreted as the tunnelling of a fluxon through the
contact. This process can be safely described within a
semiclassical approximation and we use the approach developed in
Refs.~\onlinecite{ustMQT,col} 
 to calculate the escape rate
$\Gamma=(\omega_P/2\pi) \sqrt{120\pi B}\exp(-B)$ of a fluxon
through the potential barrier. Here, $\omega_P$ is the oscillation
frequency of a fluxon near the effective potential minimum, and
$B=\int_{-\infty}^{\infty} d\tau\; {\cal L}(\tau=it)$ is described
by the Lagrangian~\eqref{l} with the classical fields determined
by Eqs.~\eqref{din}, if we add the term $j=J/J_c$ in the
right-hand-side of the first equation. In the limit $\gamma^2\gg
1$, the equation for $\varphi$ is reduced to standard coupled
sine-Gordon equations~\cite{equat}, which in the continuous limit,
$k_ys\ll 1$ and $y=ns/\lambda_{ab}$, reads $
\left(1-\partial^2/\partial
y^2\right)\left[\ddot{\varphi}+\sin\varphi\right]-\partial^2\varphi/\partial
x^2=j$. We seek a solution of the last equation in the form
$\varphi=\psi(x,y,t)+\arcsin(j)$, where the field $\psi$ obeys
\begin{equation}\label{psi}
\left(1-\frac{\partial^2}{\partial
y^2}\right)\left[\ddot{\psi}-j(1-\cos\psi)+\sqrt{1-j^2}\sin\psi\right]
-\frac{\partial^2\psi}{\partial x^2}=0.
\end{equation}
Following the experimental setup~\cite{ustMQT}, here we consider
the SIJJs having the size $L\gg s$ along the $y$ direction, i.e.,
the total number of contacts $N=L/s\gg 1$, and the size of the
SIJJs in the $x$ direction, $2d$, is smaller than the Josephson
length, $\lambda_J=\gamma\sqrt{s\lambda_{ab}/2}$.

We can linearize Eq.~\eqref{psi} in all junctions except one,
where the fluxon tunnels. The linearized equation can be solved by
using the Fourier transformation, $\psi=\sum_m\int \exp(-i\omega
t)\cos(k_{xm}x)\psi_m(y,\omega)d\omega/2\pi$, where
$k_{xm}=\lambda_c\pi (2m+1)/2d$. Since in the
experiment~\cite{ustMQT} the sample connects two bulk
superconductors, we can choose the phase difference to be zero at
the top ($y=L_1$) and bottom ($y=L_1-L$) layers of the sample, and
$y=0$ corresponds to the position of the fluxon tunneling. As a
result, we derive the solution of the linearized equations in the
form $\psi_m(y)=\psi_m(0)\sinh[q_m(L_1-y)]/\sinh[q_mL_1]$, for
$y>0$, and
$\psi_m(y)=\psi_m(0)\sinh[q_m(L-L_1+y)]/\sinh[q_m(L-L_1)]$ for
$y<0$. Here,
$q_m^2=(k_{xm}^2+\sqrt{1-j^2}-\omega^2)/(\sqrt{1-j^2}-\omega^2)$.
Following the method described in Ref.~\onlinecite{cher} and
requiring the continuity of both $\psi$ and the current flowing
through the central ($y=0$) contact, we obtain the nonlinear
equation for the phase difference in the junction with $y=0$:
\begin{eqnarray}\label{nonlocal}
\ddot{\psi}-j(1-\cos\psi)+\sqrt{1-j^2}\sin\psi=
\frac{\lambda_J^2}{\lambda_c^2}\int_{-\infty}^{\infty}
\frac{d\omega}{2\pi}e^{-i\omega t}\nonumber
\\\times\sum_m\frac{k^2_{xm}}{q_m}\frac{\sinh(q_mL_1)\sinh(q_m(L-L_1))}{\sinh(q_mL)}
\cos(k_{xm}x)\psi_m(\omega).
\end{eqnarray}
For an infinite ($L,d\rightarrow \infty$) sample, this equation
coincides with the nonlocal equation for the Josephson vortex in
the SIJJs~\cite{cher}. Eq.~\eqref{nonlocal} can be used to
describe the tunneling of a fluxon through a SIJJs with any width
$d$ and any number of layers $N$. However, such a treatment can
only be done numerically.

\begin{figure}[!htp]
\begin{center}\label{tunnel}
\includegraphics*[width=8.7cm]{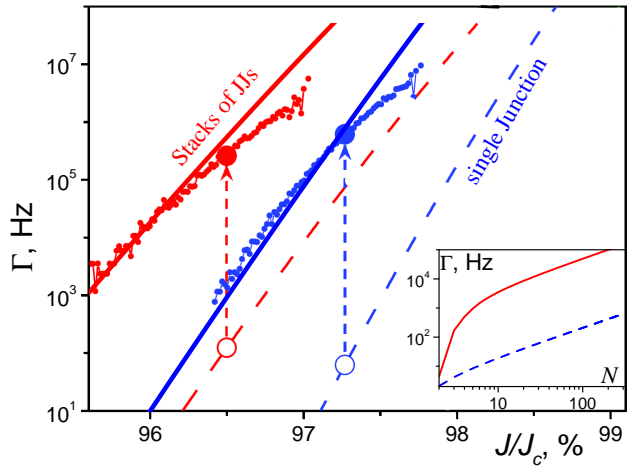}
\vspace{6cm} \caption{(Color online) The MQT escape rate $\Gamma$
versus dimensionless external current $j$. Red and blue points are
the experimental data, from Ref.~\onlinecite{ustMQT}, for two
different samples.  Red and blue dashed lines are the curves
$\Gamma_0(j)$, Eq.~\eqref{Ust}, for these samples taken
from~\cite{ustMQT}. Red and blue solid lines are the functions
$\Gamma(j)$ calculated from Eq.~\eqref{Gamma}, using data from
Table~1 of~\cite{ustMQT}, for their samples US1 and US4,
$\gamma=600$, $s=15\;$\AA. The inset shows the dependence of the
escape rate $\Gamma$ versus the number of contacts $N$ in the
SIJJs, using parameters for sample US4 of~\cite{ustMQT} at
$j=0.97$ (red solid line). The blue dashed line shows
$N\Gamma_0$.}
\end{center}
\end{figure}

Now, we adopt Eq.~\eqref{nonlocal} for the short ($d/\lambda_J\ll
1$) SIJJs used in \cite{ustMQT}, where
$d/\lambda_J\,\approx\,2\mu{\rm m}\,/\,5\mu{\rm m}\,=\,0.4$. In
this case, the phase difference $\varphi$ changes slowly with $x$
and the main contribution to the sum in the right-hand-side
of~\eqref{nonlocal} comes from the first harmonic
$k_{xm}=\pi\lambda_c/2d$. Neglecting contributions to the
tunnelling process arising from higher-frequencies,
$\omega\geq\omega_J (1-j^2)^{1/4}$, and integrating
Eq.~\eqref{nonlocal} over $dx$, we derive for the phase difference
${\bar \psi}$, averaged over $x$, the equation: $d^2{\bar
\psi}/dt^2=-\partial V/\partial {\bar \psi}$. Here the effective
potential $V({\bar\psi})$ can be written as
\begin{equation}\label{poten}
V({\bar \psi})=j(\sin{\bar \psi}-{\bar
\psi})-\sqrt{1-j^2}(\cos{\bar \psi}-1)-g_n(j)\frac{{\bar
\psi}^2}{2}\;,
\end{equation}
where $n=L_1\lambda_{ab}/s$ labels the contact through which the
fluxon tunnels,
\begin{equation}\label{gn}
g_n(j)=\frac{2(1-j^2)^{1/2}\;Q}{\pi}\;\frac{\sinh(Qn)\sinh[Q(N-n)]}{\sinh(QN)},
\end{equation}
and $Q(j)=\pi\gamma s\left(1-j^2\right)^{1/4}/2d$. For an applied
current $J$ close to $J_c$, where tunnelling was
observed~\cite{ustMQT}, we can expand both $\cos{\bar \psi}$ and
$\sin{\bar \psi}$ and, finally, derive $V({\bar \psi})=-{\bar
\psi}^2({\bar \psi}-\psi_1)/6$, where
$\psi_1(j)=3[\sqrt{1-j^2}-g_n(j)]$. Using a semiclassical
approach~\cite{col} 
(i.e., $B=\int_0^{\psi_1}[2V({\bar
\psi})]^{1/2}d{\bar \psi}$), and taking into account that the
fluxon can tunnel through any junction of the SIJJs, we derive
(now in dimensional units)
\begin{equation}\label{Gamma}
\frac{\Gamma}{\Gamma_0}=\sum_{n=0}^N\left(1-g_n\right)^{5/4}
\exp\left\{-\frac{36\,U_0}{5\,\hbar\omega_P}\left[(1-g_n)^{5/2}-1\right]\right\},
\end{equation}
where the summation is taken over all $N$ contacts. Here, the
effective Josephson frequency is
$\omega_P(j)=\omega_J(1-j^2)^{1/4}$, the height of the potential
barrier $U_0=2E_J(1-j^2)^{3/2}/3$, the Josephson energy
$E_J=\Phi_0J_c/2\pi c$, and the escape rate $\Gamma_0(j)$ for a
single Josephson junction (see, e.g.,~\cite{ustMQT}) is given by
\begin{equation}\label{Ust}
\Gamma_0(j)=\frac{6\,\omega_P(j)}{\pi}\sqrt{\frac{6\pi
U_0(j)}{\hbar
\omega_P(j)}}\exp\left(-\frac{36\,U_0(j)}{5\,\hbar\omega_P(j)}\right),
\end{equation}
Figure 2 shows $\Gamma(j)$, which very well describes experimental
results in~\cite{ustMQT}. Some deviation between the experimental
data and the theoretical prediction at high currents is due to a
significant lowering of the potential barrier resulting in a
decrease of the accuracy of the semiclassical approximation. The
dependence of $\Gamma$ on the number $N$ of junctions is nonlinear
due to the long-range interaction between different junctions,
described by the last term in the expression~\eqref{poten} for the
effective potential. This nonlinearity is strong for relatively
small $N\lesssim N_c=d/\gamma L$ and the escape rate becomes
proportional to $N$ when the SIJJs thickness $L$ exceeds the
effective interaction length $d/\gamma$. Very different types of
MQT models in SIJJs, with no quantitative comparison with
experimental data,
are also being studied in \cite{teorMQT}. For instance, here we
consider the inductive coupling among layers, which is known to be
strong, instead of the weak capacitive coupling among layers used
in \cite{teorMQT}.

{\it Conclusions.---} We analyze the quantum effects in SIJJs. We
develop a model for quantum excitations in SIJJs using two Bosonic
fields. We also describe the interactions and thermodynamics of
these fields. Moreover, we suggest a semiclassical theory of the
fluxon quantum tunneling in SIJJs, which is in good agreement with
recent remarkable experimental observations. The obtained results
might be potentially useful for future designs  of quantum THz
devices.

We acknowledge partial support from the NSA, LPS, ARO, NSF grant
No. EIA-0130383, JSPS-RFBR 06-02-91200, RFBR 06-02-16691, MEXT
Grant-in-Aid for Young Scientists No 18740224, and an EPSRC
Advanced Research Fellowship.
%
%

\end{document}